\documentclass[runningheads]{llncs}
\usepackage[pdfstartview=XYZ,
bookmarks=true,
colorlinks=true,
linkcolor=blue,
urlcolor=blue,
citecolor=blue,
pdftex,
bookmarks=true,
linktocpage=true, 
hyperindex=true
]{hyperref}

\usepackage{orcidlink}

\usepackage{comment}
\usepackage{soul}
\usepackage{xcolor}
\usepackage{subfigure}
\usepackage{amssymb}
\usepackage{amsmath}
\usepackage{amsfonts}
\usepackage{graphicx}
\usepackage{bm}
\usepackage{xfrac}
\usepackage{hyperref}
\usepackage[acronym,shortcuts, hyperfirst=false]{glossaries}
\usepackage{balance}
\usepackage{caption}
\usepackage{bbm}
\usepackage{paralist}

\newacronym{bsc}{BSC}{Binary Symmetric Channel}
\newacronym{dmc}{DMC}{Discrete Memoryless Channel}
\newacronym{snr}{SNR}{Signal-to-Noise Ratio}
\newacronym{rf}{RF}{Radio-Frequency}
\newacronym{ask}{ASK}{Amplitude Shift Keying}
\newacronym{isp}{ISP}{Internet Service Providers}
\newacronym{mpls}{MPLS}{Multi-Protocol Label Switching}
\newacronym{ospf}{OSPF}{Open Shortest Path First}
\newacronym{wan}{WAN}{Wide Area Network}
\newacronym{vpn}{VPN}{Virtual Private Network}
\newacronym{vrf}{VRF}{Virtual Routing and Forwarding}
\newacronym{as}{AS}{Autonomous System}
\newacronym{bgp}{BGP}{Border Gateway Protocol}
\newacronym{apt}{APT}{Advanced Persistent Threat}
\newacronym{ook}{OOK}{On-Off Keying}
\newacronym{mtu}{MTU}{Maximum Transmission Unit}
\newacronym{ar}{AR}{Amplitude Ratio}
\newacronym{ber}{BER}{Bit Error Rate}
\newacronym{pam}{PAM}{Pulse Amplitude Modulation}
\newacronym{lan}{LAN}{Local Area Network}

\usepackage{cuted}
\usepackage{booktabs}

\usepackage{tabularx}
\usepackage{threeparttable} 
\usepackage{multirow} 
\usepackage{rotating} 
\usepackage{bigstrut} 
\usepackage{setspace}

\newcommand{\parag}[1]{\noindent\textbf{#1. }}

\usepackage{algpseudocode}
\usepackage[ruled,vlined,noend]{algorithm2e}
\SetKwInput{kwFailure}{Failure}

\begin{document}
\title{CONNECTION:	COvert chaNnel NEtwork attaCk Through bIt-rate mOdulatioN}
\titlerunning{CONNECTION: covert channel network attack}
%
\author{Simone Soderi\inst{1,2}\orcidlink{0000-0002-1024-9470}, Rocco De Nicola\inst{1,2}\orcidlink{0000-0003-4691-7570}}
%
\authorrunning{S. Soderi}

%
\institute{IMT School for Advanced Studies Lucca, Lucca, Italy \and
Cybersecurity National Laboratory, CINI - Roma, Italy
\email{\texttt{\{name.surname\}@imtlucca.it}\\}}
%
\maketitle              
\begin{abstract}

Covert channel networks are a well-known method for circumventing the security measures organizations put in place to protect their networks from adversarial attacks.
This paper introduces a novel method based on bit-rate modulation for implementing covert channels between devices connected over a wide area network.
This attack can be exploited to exfiltrate sensitive information from a machine (i.e., covert sender) and stealthily transfer it to a covert receiver while evading network security measures and detection systems.
We explain how to implement this threat, focusing specifically on covert channel networks and their potential security risks to network information transmission. The proposed method leverages bit-rate modulation, where a high bit rate represents a `1' and a low bit rate represents a `0', enabling covert communication.
We analyze the key metrics associated with covert channels, including robustness in the presence of legitimate traffic and other interference, bit-rate capacity, and 
bit error rate. 
Experiments demonstrate the good performance of this attack, which achieved $5$~bps with excellent robustness and a channel capacity of up to $0.9239$~$\sfrac{bps}{Hz}$ under different noise sources. Therefore, we show that bit-rate modulation effectively violates network security and compromises sensitive data.

\keywords{Covert channel  \and Network security \and Cyber range \and Bit-rate.}
\end{abstract}

\newpage
\noindent\rule{8.4cm}{1pt}\\
Please cite this version of the paper:\\

Soderi, S., De Nicola, R. CONNECTION: COvert chaNnel NEtwork attaCk Through bIt-rate mOdulatioN. In: Shao, J., Katsikas, S.K., Meng, W. (eds) Emerging Information Security and Applications. EISA 2023. Communications in Computer and Information Science, vol 2004. Springer, Singapore. https://doi.org/10.1007/978-981-99-9614-8\_11
\\
You may use the following bibtex entry:
\begin{verbatim}
@InProceedings{Soderi2023Connection,
author="Soderi, Simone
and De Nicola, Rocco",
editor="Shao, Jun
and Katsikas, Sokratis K.
and Meng, Weizhi",
title="CONNECTION: COvert chaNnel NEtwork attaCk Through bIt-rate mOdulatioN",
booktitle="Emerging Information Security and Applications",
year="2024",
publisher="Springer Nature Singapore",
address="Singapore",
pages="164--183",
isbn="978-981-99-9614-8",
doi=10.1007/978-981-99-9614-8_11
}

\end{verbatim}
\noindent\rule{8.4cm}{1pt}

%
%
\section{Introduction}\label{sec:intro}
In today's digital era, information exchange via networks is widespread and diverse, covering voice, text, images, and video~\cite{2018:bridges_towards_katz,2020:6g-horizons}. Despite its vital role in modern society, networked communication is vulnerable to security breaches. Malicious actors can exploit communication channels, conducting covert attacks to extract sensitive data or establish hidden backdoors~\cite{2017:information_hiding}. Indeed, \textit{covert network channels}, 
concealing unauthorized information transfer, present a substantial security risk. These channels can be crafted by manipulating network protocols or using steganography to hide data within innocuous information. Their primary aim is discreet information transfer, evading detection and network disruption. This stealthy nature makes covert channels a potent tool for attackers, allowing them to access data secretly.
Examining covert network channels is vital for network security. Enhancing detection and mitigation strategies is crucial to prevent data leaks and unauthorized access. Covert channels pose tangible threats to networks and data processes~\cite{2015:info_hiding_malware_detection}. Their goal is hidden infiltration, increasingly challenging in evolving security landscapes. Incorporating covert channel detection tools in network defence is essential.
Data can be surreptitiously transferred via various methods, often exploiting well-known internet protocols~\cite{2018:new_trends_info_hiding}. 
Monitoring covert channel actions can prevent future attacks. This paper introduces a novel technique relying on bit-rate modulation for covert communication between networked devices. The approach aims to evade detection systems by exploiting bit-rate variations. To the best of our knowledge, this mechanism is new and calls for further research to curb its impact.

\parag{Contribution} The main contribution of this paper is the design of a new covert channel to attack enterprise networks under different noise sources. The basic intuition is that the attacker is not interested in intercepting the data transiting the network; rather, she/he wants to control the bit-rate at which it is transmitted. By modulating the bit-rate appropriately, the attacker can associate some pieces of information with a high bit-rate and others with a low bit-rate. Although simple, this technique is new in the covert channel scenario. Other contributions of this paper include the \textit{artefact} implementation of the covert sender and receiver and the description of experiments on the performance of the proposed attack in a cyber range that emulates an enterprise data network.

The rest of the paper is organized as follows. Section~\ref{sec:background} describes the main background concepts used in this work, while Section~\ref{sec:related} contains a short overview of revising the literature about covert channels. Section~\ref{sec:attacker-model} introduces the attacker model, while Section~\ref{sec:modulation-covertch} describes the actual implementation of the covert sender and receiver. Section~\ref{sec:eval} presents our experiments and its results. Finally, Section~\ref{sec:conclusions} concludes the paper.
\section{Background}\label{sec:background}
This section provides the basic concepts for understanding covert channels and how we exploited them during our experiments.
{\color{black}Furthermore, it provides the basic concepts of network architectures that span wide geographical areas, which will be used in an emulated environment (a \textit{cyber-range}) to assess our attack model.}

\subsection{Covert Channel Characteristics} \label{sec:covert-channel-charact}
The creation, maintenance, and effective operation of covert channels depend on balancing fundamental characteristics: anonymity, robustness, bit error rate, and channel capacity.
These prerequisites collectively serve as crucial factors in assessing the effectiveness of a covert channel, as they determine its ability to transmit information while eluding network monitoring security systems~\cite{2022:covert_ch_detection}. The paramount consideration in crafting a successful covert channel is skillfully achieving and harmonizing all these essential features.

\parag{Anonymity} This feature, often synonymous with unobservability, is essential in establishing the covert nature of the channel. To ensure evasion from potential detection mechanisms, the communication via a covert channel must adeptly mimic ordinary network traffic. This requirement extends beyond simple masquerading; the channel's data patterns and behavioural characteristics must not show statistically significant deviations compared to normal traffic patterns. Additionally, the channel must remain immune to various intrusion detection techniques, including anomaly-based and signature-based detection systems.

\parag{Robustness} This attribute of a covert channel represents its resilience to alterations, noise, and data losses. Covert channels should exhibit high data integrity and guarantee accurate and reliable data transmission despite potential disruptions. For instance, channels must be resilient against network jitter, packet loss, or intentional disruptions. Furthermore, the channel should be robust against varying network conditions, such as changes in network load or transmission rates.

\parag{\ac{ber}} The bit error rate measures the frequency of transmission errors or bit errors during communication. It is crucial to evaluate the error rate of the covert channel as it can affect the reliability and accuracy of the transmitted data. Higher error rates can lead to information loss or corruption, compromising covert communication.

\parag{Coexistence with Legitimate Traffic} Analyzing the behaviour of a covert channel when it coexists with legitimate network traffic is essential. This investigation helps uncover potential interference or congestion issues that may arise when covert communication overlaps with regular network traffic. Understanding these coexistence characteristics is instrumental in designing covert channels that are not only efficient but also highly stealthy.

\begin{figure}[t]
    \centering
    \includegraphics[width=0.9\columnwidth]{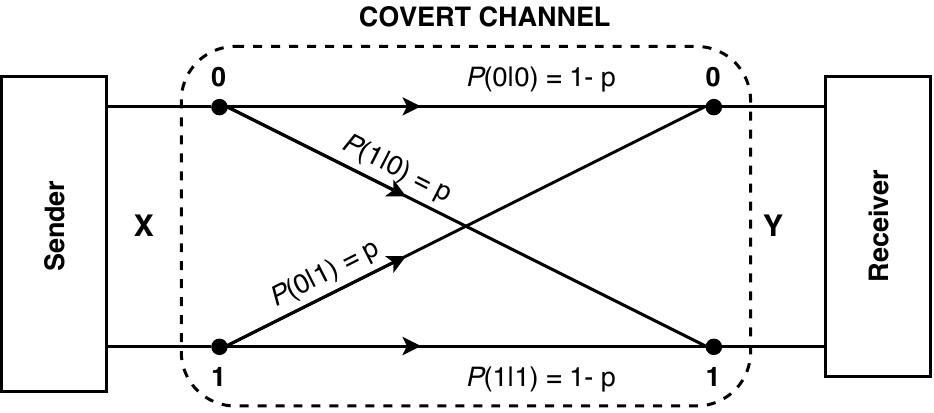}
    \caption{Binary Symmetric Channel model.}
    \label{fig:bsc}
\end{figure}

\parag{Channel Capacity}
When designing a covert communication channel using unintended mechanisms, understanding its capacity is crucial. This capacity signifies the maximum reliable information rate transmitted, a vital metric for assessing efficiency.
We use the \ac{bsc} model~\cite{Proakis2007digital}, shown in Figure~\ref{fig:bsc}, a discrete variant of the \ac{dmc} model. It has binary input/output symbols, $X, Y \in {0,1}$, and accounts for noise-causing errors ($p$).
For instance, a `0' or `1' could be sent, with $p$ probability of receiving the opposite due to noise ($P(1|0)$ or $P(0|1)$). Conversely, $(1-p)$ is the probability of correct reception ($P(0|0)$ or $P(1|1)$). This models symmetric errors.In this case, the channel capacity, $C$, 
can be mathematically determined using the following equation~\cite{Proakis2007digital}. It quantifies optimal error-free data transfer while optimizing efficiency and minimizing detection risk and is expressed in~$\sfrac{bps}{Hz}$.
%
%
%
\begin{align}
    C =  \max_{P} \mathbb{I}(X;Y) = 1 - H(p)  
      =   1 - p \log_2\left(\frac{1}{p}\right) - (1-p) \log_2\left(\frac{1}{1-p}\right), \label{EQ:CH_C}
\end{align}
where $\mathbb{I}(X;Y)$ is the mutual information between the input $X$ and output $Y$, $H(p)$ is the binary entropy function, and $p$ is the error probability.

\subsection{Brief Overview of Enterprise Networks} \label{sec:enterpr_net}

\begin{figure}[t]
    \centering
    \includegraphics[width=0.85\columnwidth]{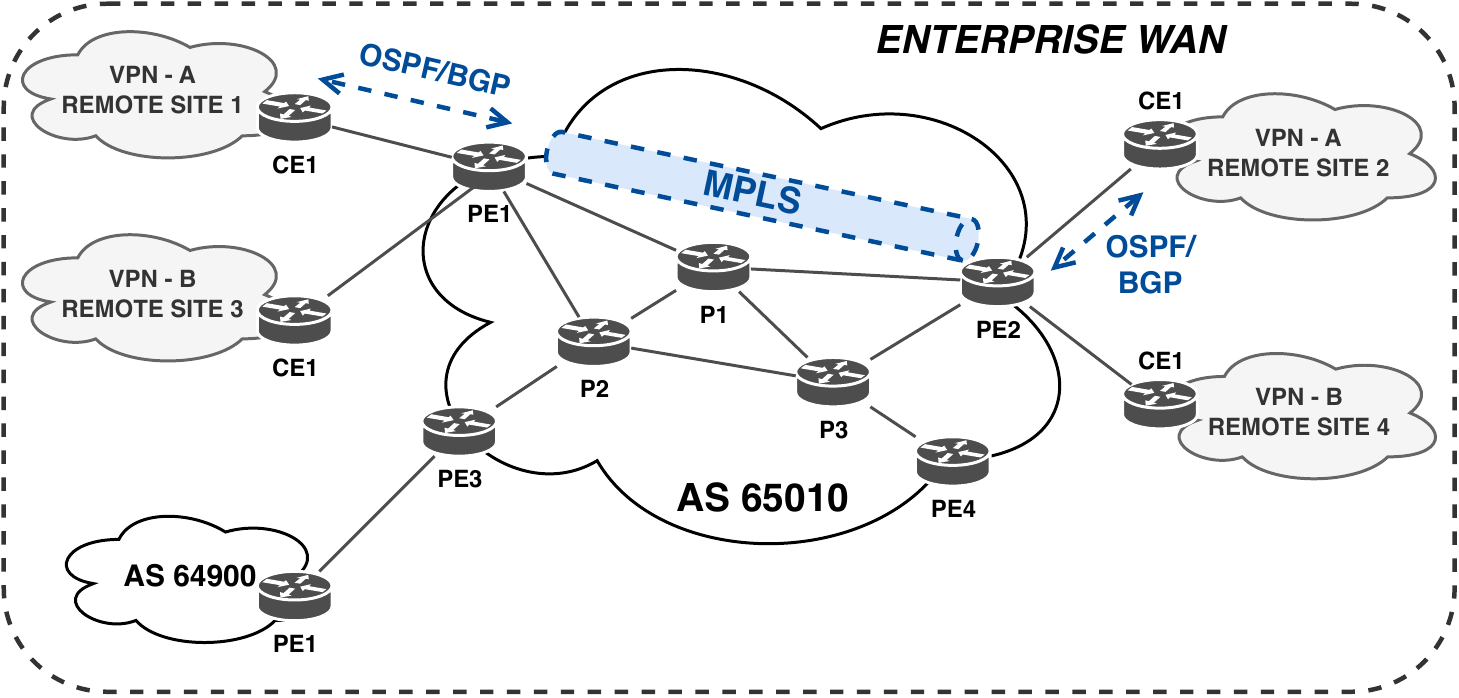}
    \caption{Overview of service and enterprise networks.}
    \label{fig:wan}
\end{figure}

In computer networks and communication infrastructures, service and enterprise networks emerge as two distinct interconnected types~\cite{CISCO2007MPLSDCN}, with the \ac{mpls} protocol playing a central role. \ac{mpls} governs large \acp{isp} networks and enterprise network backbones, utilizing various routing protocols like \ac{ospf}, internal Border Gateway Protocol (iBGP), and \ac{bgp} between different \acp{as}~\cite{DeGhein2016MPLS,CISCO2020MPLSVPN}. \ac{mpls} employs labelling mechanisms for efficient packet routing, ensuring rapid and deterministic switching.
For example, consider routing an IP packet through a Layer 3 Virtual Private Network (L3VPN) from a CE device at \textit{remote Site~1} to \textit{remote Site~2}, as depicted in Figure~\ref{fig:wan}. The packet is labelled at the ingress PE device, dictating its path. P routers forward the packet using its label, and the label is removed at the egress PE device before reaching the destination CE device at \textit{Site~2}. This process is controlled by \ac{vrf} instances at the PE devices, segregating network traffic based on \ac{vpn} membership.
\ac{mpls} plays a crucial role in modern network architecture by efficiently routing data packets using labels. This approach facilitates streamlined routing across remote sites within networks, avoiding the complexities of destination-based routing algorithms. With \ac{vrf} instances, \ac{mpls} ensures secure and efficient connectivity between different network sites, highlighting its pivotal contribution to managing data flows within intricate network structures~\cite{behringer2005mpls}.
%

\section{Related Works}\label{sec:related}
In 1973, Lampson defined the \textit{covert channel} as a communication channel not intended for information transfer~\cite{1973:lampson_covert_ch}. Such type of communication falls within a more general discipline of data manipulation known as \textit{information hiding} or \textit{data hiding}.  Nowadays, covert channels are carefully engineered to evade detection and circumvent network security policies by establishing new communication paths, causing the disruption of security mechanisms and making their creation and maintenance challenging~\cite{2022:etherled-covert-channel,2022:soderi-covert-channel-fl,2022:TCAN_conti}.
Since covert channels exploit unconventional aspects, it is easy to imagine that there are different types of them. It is, therefore, important to mention a couple of them.

\parag{Covert Timing Channels} These covert channels transmit information based on the packet arrival times rather than the actual packet content~\cite{2013:timing_covert_channel_retx,2014:PHY_covert_ch,2011:timinig-wm-nordsec}. They can be subdivided into two types: \textit{timing} and \textit{sorting} channels. In the first, the sender uses predefined time intervals to convey information; for instance, the reception of a packet could represent a binary `1', and the absence of a packet could represent a `0'. More sophisticated timing channels might use inter-packet arrival times to convey data. In such a scenario, a long delay may signify a `0' while a short delay indicates a `1'. Whereas in sorting channels, the sequence of packet arrivals forms the transmitted message.
Given that such channels secretly exploit temporal data to forward information, they face significant challenges, such as synchronization between the sender and receiver, especially when the network introduces jitters or noise. However, despite their relatively low throughput, the high anonymity makes them efficient enough for covert communication. 
Multiple techniques aim to enhance statistical detection's resilience, such as mimicking genuine traffic patterns or employing randomized inter-packet delays~\cite{2008:tcp_covert_timing_channel}.
    
\parag{Covert Network/Storage Channels} These channels manipulate different layers of the Internet Protocol stack to facilitate covert communication. The sender, who can control parts of the network stack, modifies protocol headers, fragments, checksum values, or packet transmission timing to conceal the information~\cite{2009:storage_covert_ch_fragments,2015:network-covert-ch-nordsec}. The hidden messages could be embedded in unused bits or fields of protocol headers, such as the IP identification field, the IP fragment offset, the TCP sequence number field, and TCP timestamps~\cite{2022:network_iot_covert_channel,2009:stega_storage_covert_channel}. While embedding messages in packet headers provides an easy way to create a covert channel, these channels are equally simple to detect and prevent. An adversary can monitor specific fields of packet headers for detection or sanitize these fields for prevention. Furthermore, it is worth noting that payload embedding is generally avoided in the spirit of maintaining covert communication, as this could reveal the existence of the covert channel. 
For instance, a spread-spectrum watermarking technique can track data covertly flows through a network~\cite{2009:detection-SS-wm-nordsec}.

Covert channels, characterized as \textit{policy-violating} and unconventional stealthy communication mediums in a system's design, are habitually employed to secretly transfer sensitive information, potentially for data exfiltration or malware communications. Moreover, these channels can serve as vehicles for circumventing cybersecurity policy in companies' perimeter. Covert channels have been observed in many environments, ranging from networks, cyber-physical systems~\cite{2021:cyberphy_covert_ch}, and local processes/systems to out-of-band scenarios like ultrasonic sound~\cite{2018:guri_MOSQUITO_cpvert_ch}, light~\cite{2022:etherled-covert-channel}, power supply~\cite{2023:guri_powersupply_covert_ch}, radio frequency~\cite{2022:covert_ch_LoRa,2014:soderi_fingerprint}, magnetic fields~\cite{2021:guri_MAGNETO_covert_channel}, or temperature~\cite{2019:guri_HOTSPOT,2015:guri_thermal_covert_ch}.
Overall, each type of covert channel has advantages and drawbacks, and the choice between them often depends on the specific requirements of covert communication and the constraints imposed by the network environment.
Furthermore, a prevalent characteristic among covert channels is the low data transfer rate, exploiting the potential of protracted hiding. Covert channels achieving speeds of $100$~bps are classified as fast in this context~\cite{2012:high-speed_covert-ch}.

In addition to timing and storage channels, other types of covert channels are worth mentioning, even if they are less important in the context of this paper:  
\begin{itemize}

    \item[($i$)] \textit{covert thermal channels} use a process to modulate the CPU usage and change the machine's temperature, which is then sensed by a thermal sensor controlled by another process~\cite{2015:guri_thermal_covert_ch}; 
    \item[($ii$)] \textit{covert acoustic channels} leverage the audible or ultrasonic sound spectrum for data exfiltration~\cite{2020:guri_acoustic_covert_ch,2018:guri_MOSQUITO_cpvert_ch}.
    \end{itemize}

\section{Attacker Model}\label{sec:attacker-model}
In this section, we illustrate the proposed model of our attacker. The attacker's main objective is to illegally establish a communication conduit, herein termed a \textit{covert channel} between two specific entities: the \textit{sender} and the \textit{receiver}. They must be intended as equipment interconnected via a WAN, as described in Section~\ref{sec:enterpr_net}.
In the assumptions of our adversarial model, both the sender and the receiver fall under the attacker's control. 
To illustrate, consider the situation wherein the sender and the receiver are devices that malware deployed by the attacker has compromised. This malware can execute arbitrary code on the infected devices, creating a
\textit{connection}.
{\color{black}It is important to emphasize that when constructing our attack model for establishing covert channels, we control and integrate features of both senders and receivers, exactly like in many other approaches to the same issue.}
Many studies focusing on covert channels have incorporated an attack model that, like ours, is characterized by a transmitter-receiver structure in which the attacker requires the same control capabilities. Some examples from recent literature can be found in these references  \cite{2022:maritime-covert-channel-esorics,2022:soderi-covert-channel-fl,2022:etherled-covert-channel,2019:router-covert-channel-usenix,2022:nf-covert-channel}, but none of the considered approaches relies on bit-rate modulation.
\begin{figure}[t]
    \centering
    \includegraphics[width=0.95\columnwidth]{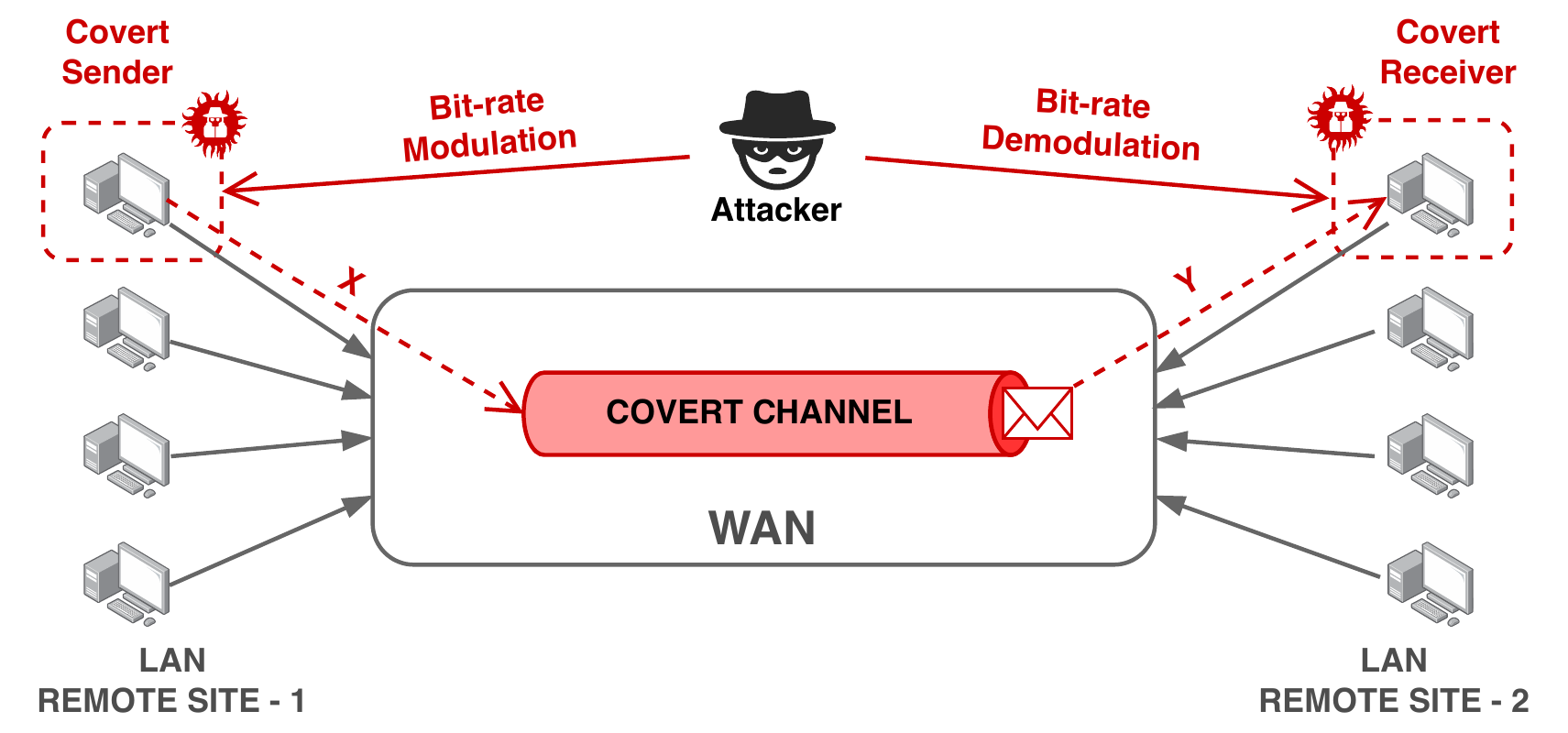}
    \caption{Attack model where the WAN can be an enterprise network or the Internet. Sender and Receiver are two devices connected through the WAN.}
    \label{fig:attacker-model}
\end{figure}

The concept of an \ac{apt} delineates an attack campaign whereby an adversarial entity maintains an unauthorized presence within a targeted network, thereby gaining high-sensitivity access to data~\cite{2016:APT}. The attacker model we propose, named CONNECTION - “COvert chaNnel NEtwork attaCk Through bIt-rate mOdulatioN", involves both sender and receiver elements, as illustrated in Figure~\ref{fig:attacker-model}. 
This attack falls into the category of a covert timing channel. The compromise of the two entities can be achieved through various methods, including but not limited to supply chain attacks, the use of social engineering techniques, or the exploitation of hardware with pre-installed software or firmware. The attacker does not require special permissions to carry out this attack once the machines have been infected. It's worth noting that even highly secure facilities, including air-gapped ones, are not immune to such exploits, as demonstrated in recent years, with Stuxnet~\cite{2013:real-story-stuxnet} being one of the most famous examples

After successfully infecting the two designated devices, the malicious software initiates a process to identify the communication partners of the sender. In this scenario, the attacker may not need to know the exact address of the sender; the target network's address range or the IP address subnet could be sufficient.

Under the attacker's control, the sender initiates covert communication attempts with each IP address within the target network's address space. Upon successfully detecting this covert communication, the (infected) receiver
responds to the sender by modulating its bit-rate.
To be more precise, when the sender selects a specific IP address, it begins transmitting a bitstream (represented as $X$ in Figure~\ref{fig:attacker-model}), using bit-rate variation as the communication method: a high bit-rate indicates the transmission of a '1' bit, while a low bit-rate indicates the transmission of a '0' bit.

{\color{black}Conversely, the receiver using a traffic dump can demodulate the encoded information, resulting in receiving the message $Y$. By observing the traffic shape,  often referred to as the \textit{envelope}, exchanged between these two devices,  an amplitude modulation such as a \ac{pam} or an \ac{ook} modulations~\cite{Proakis2007digital} could be identified. However, in our case, this is achieved by controlling the \textit{network throughput}.
Implementing this type of covert channel depends on the specifications of the victim device. The following section will discuss how to carry out this attack. }

\section{Bit-rate modulation as Covert Channel}\label{sec:modulation-covertch}

Let us, therefore, consider the context described in Section~\ref{sec:attacker-model}, i.e., a threat actor operating on two devices communicating over a WAN (see Figure~\ref{fig:wan}). 
Given its extensive use, at this time, we will use the MITRE ATT\&CK framework~\cite{2018:strom_mitre} as a reference to frame the activities of the attacker. In particular, he/she must plan tactics and use techniques to implement the covert channel.
There will be a pre-attack phase in which the attacker will conduct a \textit{reconnaissance} phase to acquire useful information to \textit{weaponize} and adapt the malware that will create the covert channel between the sender in \textit{Site~1} and the receiver in \textit{Site~2}.
The techniques needed to effectively realize this covert communication, and in particular, how the covert channel between  the \textit{sender} and the \textit{receiver} is established, are explained in the following.

\subsection{Covert Sender} \label{sec:sender}
Algorithm~\ref{alg:sender} outlines a procedure for transmitting a $bitstream$ over a covert network. The process employs the principles of bit-rate modulation, effectively performing an amplitude modulation (e.g., \ac{pam}) of the throughput. This technique ingeniously hides the communication, making it robust against conventional detection strategies.

The procedure begins with identifying the essential elements required to execute the algorithm, which includes the amplitude modulation parameters $b1$ and $b0$, i.e., the high bit-rate levels to transmit $1$ and the low bit-rate levels to transmit $0$. 
The only information the sender must have is the IP address (or the addressing of the destination LAN) and port of the receiver. 
An attacker can potentially obtain this information in the reconnaissance phase through social engineering operations such as phishing or other means. It is, however, assumed that the sender and receiver are already connected via the WAN.

Upon initiation, the algorithm processes each bit in the $bitstream$. Depending on whether the current bit is `1' or `0', the payload is generated using the modulation parameters $b1$ or $b0$, respectively. This operation is performed by the $GeneratePayload$ function, which generates random data at the corresponding bit rate, i.e., $b0$ or $b1$.
Once the payload is generated, the algorithm encapsulates it into a network UDP packet using the $ConstructPacket$ function. This packet, intended for the receiver, contains the payload and is routed to the specified receiver's IP and port number. 
We have used the UDP protocol without losing generality to maximize control over the timing of the data sent.

To comply with \ac{mtu} constraints, specifically in Ethernet networks, and to avoid unintended fragmentation at the network layer, the constructed packet is fragmented into smaller pieces of a predetermined size, which here is set at $1472$~bytes. The  fragmentation is implemented using the $FragmentPacket$ function. After the fragmentation, each fragment is sent in a sequence using the $TransmitFragment$ function. This phase represents the transportation of each bit in the bitstream as an independent network packet.
In essence, this algorithm executes the covert transmission of the bitstream over the WAN. Applying the bit-rate modulation mechanism modulates the amplitude of the network throughput, such as an \ac{pam} signal modulation.
This mechanism, combined with the fact that the algorithm does not modify the content of the packets, makes the transmission highly resistant to detection, thereby improving the stealth and reliability of the covert channel.

\begin{algorithm}[t]
\setstretch{0.9}
\footnotesize
\SetAlgoLined
\DontPrintSemicolon
\KwIn{$b1, b0, bitstream, receiverIP, receiverPORT$}
\KwOut{Transmit $bitstream$ covertly over the network}

\SetKwFunction{FMain}{Main}
\SetKwFunction{FGeneratePayload}{GeneratePayload}
\SetKwFunction{FConstructPacket}{ConstructPacket}

\SetKwProg{Fn}{Function}{:}{}

\Fn{\FGeneratePayload{$b$}}{
    $payload$ $\gets$ empty byte array\;
    \For{$i \gets 1$ \KwTo $b$}{
        $rand\_byte$ $\gets$ random integer from 0 to 255\;
        append $rand\_byte$ to $payload$\;
    }
    \KwRet{$payload$}\;
}

\Fn{\FConstructPacket{$receiverIP, receiverPORT, payload$}}{
    $pkt$ $\gets$ Ether / IP(dst = $receiverIP$) / UDP(dport = $receiverPORT$) / Raw($payload$)\;
    \KwRet{$pkt$}\;
}

\BlankLine
\Fn{\FMain{}}{
\For{$bit \in bitstream$}{
    \uIf{$bit = 1$}{
        $payload$ $\gets$ \FGeneratePayload($b1$)\;
    }
    \Else{
        $payload$ $\gets$ \FGeneratePayload($b0$)\;
    }
    $pkt$ $\gets$ \FConstructPacket($receiverIP, receiverPORT, payload$)\;
    $frags$ $\gets$ FragmentPacket$(pkt, fragSize$)\; 
    
    \For{$frag \in frags$}{
        TransmitFragment($frag$)\;
    }
}
}
\textbf{End}

\caption{Sender's Transmission  using Bit-rate Modulation}
\label{alg:sender}
\end{algorithm}

\subsection{Covert Receiver}
In the covert communication scenario, the receiver starts by passively monitoring or sniffing UDP network traffic on a specific port, e.g., using the network traffic capture utility, \textit{tcpdump}~\cite{tcpdump}. 
It saves it in a PCAP file for further processing, i.e., demodulation.

\begin{algorithm}[!h]
\setstretch{0.9}
\footnotesize
\SetAlgoLined
\DontPrintSemicolon
\KwIn{$PCAP\_FILE, receiverIP, receiverPORT, PREAMBLE$}
\KwOut{Demodulated bitstream and bitstream synchronization}

\BlankLine

\SetKwFunction{FMain}{Main}
\SetKwFunction{FFilterPacket}{FilterPacket}
\SetKwFunction{FDemodulatePacket}{DemodulatePacket}
\SetKwFunction{FSyncCommunication}{SyncCommunication}

\SetKwProg{Fn}{Function}{:}{}

\Fn{\FFilterPacket{$pkt$, $receiverIP$}}{
    \KwIn{$pkt$ (packets from the PCAP file)}
    \KwOut{Packets considered for further processing}
    
    $fpkts$ $\gets$ Ether in pkt and IP in pkt and pkt[IP].src == receiverIP and pkt[IP].proto == 17
    
    \Return $fpkts$
}
\BlankLine
\Fn{\FDemodulatePacket{$all\_packets\_counts$}}{
    \KwIn{$all\_packets\_counts$ (a list of packets with the same id)}
    \KwOut{Demodulated bitstream}
    $threshold$ $\gets$ 0.8 * max($all\_packets\_counts$)\;
    $demodulated$ $\gets$ Empty list\;
    \For{rate in $all\_packets\_counts$}{
        \uIf{rate $\geq$ threshold}{
            Append $1$ to $demodulated$\;
        }
        \Else{
            Append $0$ to $demodulated$\;
        }
    }
    \Return $demodulated$
}
\BlankLine
\Fn{\FSyncCommunication{$demodulated, PREAMBLE$}}{
    \KwIn{$demodulated$ (the bitstream), $PREAMBLE$}
    \KwOut{$start\_bitstream\_index$ (after the preamble)}
    $preamble\_index$ $\gets$ SynchPreamble($demodulated, PREAMBLE$)\;
    \If{$preamble\_index is not empty$}{
        $start\_bitstream\_index$ $\gets$ last element of preamble\_index + 1\;
    }
    \Return $start\_bitstream\_index$
}
\BlankLine
\Fn{\FMain{}}{
    $packets$ $\gets$ ReadCaptureFile(PCAP\_FILE)\;
    $filtered\_packets$ $\gets$ \FFilterPacket($packets, receiverIP$)\;
    $all\_packets\_counts$ $\gets$ CountPacketsByIdOnPort($filtered\_packets$, $receiverPORT$)\;
    $demodulated$ $\gets$ \FDemodulatePacket{$all\_packets\_counts$}\;
   $start\_bitstream\_index$ $\gets$ \FSyncCommunication{$demodulated, PREAMBLE$}\;
    \Return $demodulated, start\_bitstream\_index$
    }
    \textbf{End}
\caption{Covert receiver's demodulation and synchronization}
\label{alg:receiver}
\end{algorithm}

The receiver analyses the network traffic it has just captured following the procedure in Algorithm~\ref{alg:receiver}. It filters only those packets that might contain secret information, demodulates the bits, and synchronizes to a known \textit{preamble}. The preamble needed to synchronise the communication is also sent using bit-rate modulation. Thus, after acquiring incoming packets, the receiver's algorithm applies a filter ($FilterPacket$) to discard all packets irrelevant for transmission, retaining only those received
that use UDP for transport (indicated by protocol number 17).

As mentioned earlier, during transmission, the sender can employ packet fragmentation. 
When transmitting a `1', which uses a high bit-rate, it will generate numerous fragments, while when transmitting a `0', it will use a much lower bit-rate and, therefore, fewer fragments or even no fragments if the payload is shorter than the \ac{mtu}. 
After the filtering process, it becomes necessary to group the packets based on their identification (ID) and count the number of packets in each group arriving at a specific port 
The resulting count serves as an indicator for the bit-rate, thus acting as an implicit representation of the value of the bit originally transmitted. The $DemodulatePacket$ function analyzes this counter, and the higher the packet count, the more likely it is that the corresponding bit value in the original bit stream was a `1' and vice versa. To translate these counts into a demodulated bit stream, a threshold comparator is used, and then a threshold value is defined, for example, $80$\% of the maximum observed count. Counts that meet or exceed this threshold are interpreted as `1', while those that fall below are interpreted as `0'.
The demodulated bit stream is scanned for this sequence, and the index immediately following the end of the sequence is taken as the starting point for the actual data in the bit stream.

{\color{black}This algorithm provides a robust and ingenious approach to covert communication by exploiting the underlying characteristics of network transmissions and their inherent variability in the number of packets to encode and transmit data secretly.}
{\color{black}Though this method operates with low throughput, the most significant advantage is the possibility that it offers a method for exfiltrating valuable information from within an organization. For example, one could send externally the file $/etc/passwd$, which in a few bytes (typically $<100$~B) contains the accounts and associated passwords of Unix computers.}

\section{Evaluation of the Covert Channel} \label{sec:eval}

\begin{figure}[t]
    \centering
    \includegraphics[width=0.9\columnwidth]{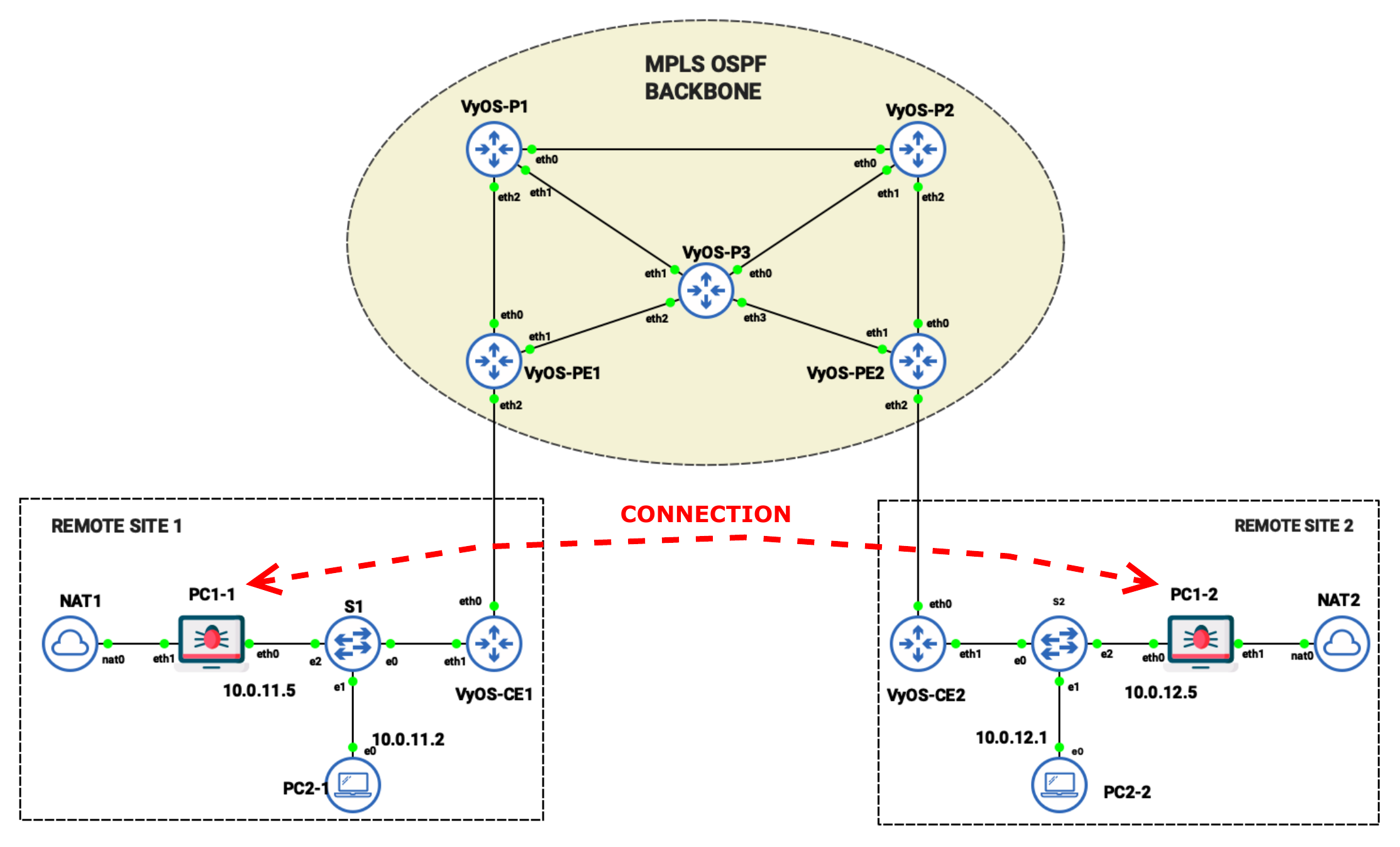}
    \caption{Covert channel (CONNECTION) deployed in a cyber-range.}
    \label{fig:cyber-range}
\end{figure}

In this section, we will evaluate the usefulness of the implemented covert channel, which uses bit-rate modulation, in a {\color{black}virtual} scenario (see Figure~\ref{fig:cyber-range}). 
This evaluation aims to understand the potential risks associated with the covert channel and assess its effectiveness as a means of unauthorized communication.
For our experiments, we relied on a controlled environment known as a cyber range, where we set up a secure network infrastructure comprising an MPLS backbone connecting two remote sites.
The cyber range provides a controlled environment to evaluate the covert channel's utility while considering achievable bit-rate, \ac{ber},  robustness, and coexistence with legitimate traffic.

We deployed the attack model sketched in Figure~\ref{fig:attacker-model} in the cyber range  using the Graphical Network Simulator-3 (GNS3)~\cite{gns3}. This network software emulator combines virtual and real devices to simulate complex networks. In this virtual environment, we established an MPLS backbone based on OSPF and BGP protocols. The backbone consisted of three routers P, two PE, and two CE, all running on the VyOS open-source network operating system~\cite{vyos}.
The cyber range connected two remote sites, each with an infected machine as the covert sender (PC1-1) and the covert receiver (PC1-2). The aim is to use the malicious covert channel to exfiltrate sensitive information from PC1-1.
To achieve this, the covert sender implements Algorithm 1 using a Python script that uses \textit{Scapy}~\cite{scapy}, a Python program for network packet manipulation. Scapy facilitated the generation, modulation, and transmission of covert payloads.
The covert receiver implements Algorithm~\ref{alg:receiver} using tcpdump, Python, and Scapy. Tcpdump allowed real-time network traffic capture, while Python and Scapy enabled the demodulation and analysis of received packets. Algorithm~\ref{alg:receiver} decoded the covert bitstream, recovering the transmitted information.

\subsection{Results of Adversary Emulation in the cyber range}\label{sec:results}

\begin{table}[b]
    \small
	\centering
	\captionsetup{justification=centering}
	\caption{Scenario parameters for the covert channel experiments.} \label{TAB:PARAM}
	\begin{threeparttable}
		\renewcommand{\arraystretch}{1}  
		\begin{tabularx}{\textwidth}{p{0.7 \textwidth} p{0.3 \textwidth}}
			\hline\noalign{\smallskip}
			\bfseries Parameter & \bfseries Value \\
			\hline\noalign{\smallskip}
		
			Modulation type  &  \ac{pam} \\
			
			Amplitude ratio ($AR$)  & $6$, $4.5$, $3$, $2$\\
   
            Bitstream (with $16$ bits of preamble) & $48 \div 144$~bits \\
			 
			Jitter & ($5$, $10$)~$ms$ \\
						
			Delay &  ($10$, $20$)~$ms$ \\
			
			Packets loss & ($1, 2, 5, 10, 15$)\% \\
			
			Packets corruption  & ($10, 20$)\% \\
			
			Frequency dropping threshold &     ($21, 51, 101, 201, 301$) \\
				
			\hline\noalign{\smallskip}
		\end{tabularx}
	\end{threeparttable}
\end{table}

We conducted a comprehensive set of experiments 
to evaluate the performance of the covert channel. The experiments considered the metrics outlined in Section~\ref{sec:covert-channel-charact}, providing valuable insights into the channel's behaviour and capabilities. The parameters and details of these tests are summarized in Table~\ref{TAB:PARAM}.
In particular, we investigated the influence of the \ac{ar} in the bit-rate modulation technique. The \ac{ar} represents the ratio between the number of bytes used to transmit bit `1' ($b1$) and the number used to transmit bit `0' ($b0$). For instance, if we allocated $6000$ bytes for bit `1' and $1000$ bytes for bit `0', the resulting \ac{ar} would be $6$.
To capture a comprehensive perspective, we collected statistical data on transmitting multiple bitstreams with varying lengths ranging from $48$ to $144$ bits (including the $16$ preamble bits). We computed the mean and standard deviation of the bit-rate and \ac{ber} by analyzing these measurements. These statistical measures allowed us to gain deeper insights into the channel's performance and robustness.
%
\begin{figure}[t]
    \centering
    \includegraphics[width=0.9\columnwidth]{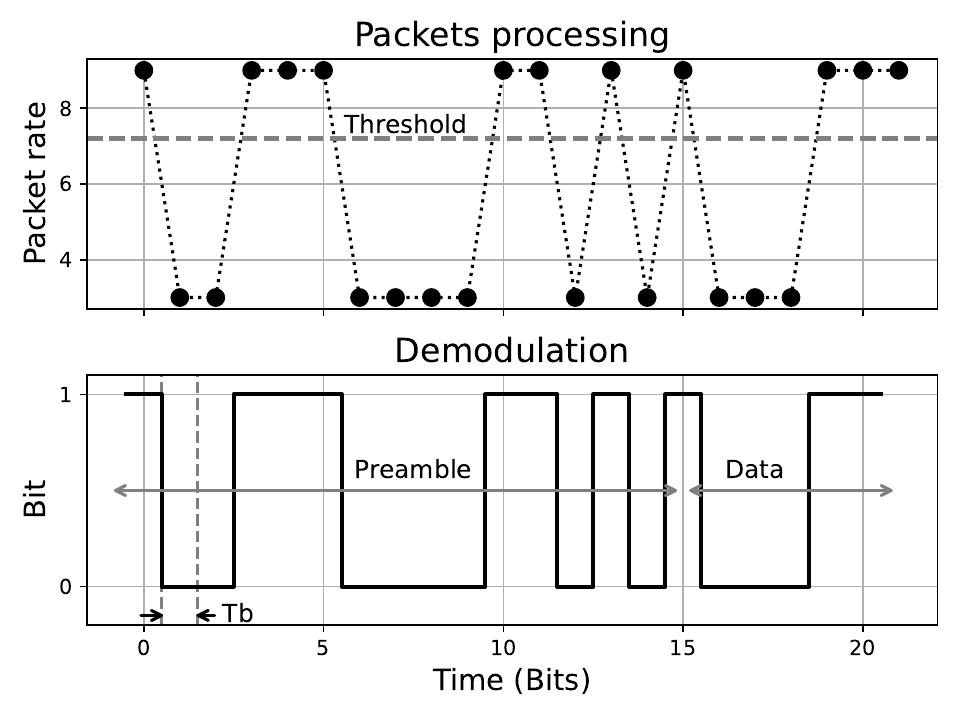}
    \caption{Transmission of a bitstream (preamble and data) over the covert channel.}
    \label{fig:tx}
\end{figure}

Figure~\ref{fig:tx} shows the covert channel in action when transmitting a short bitstream (i.e., $16$ bits of preamble and $6$ bits of data) and how demodulation (Algorithm~\ref{alg:receiver}) is done by assessing whether the packet rate is above or below the threshold.

\begin{figure}[!t]
    \centering
    \includegraphics[width=0.9\columnwidth]{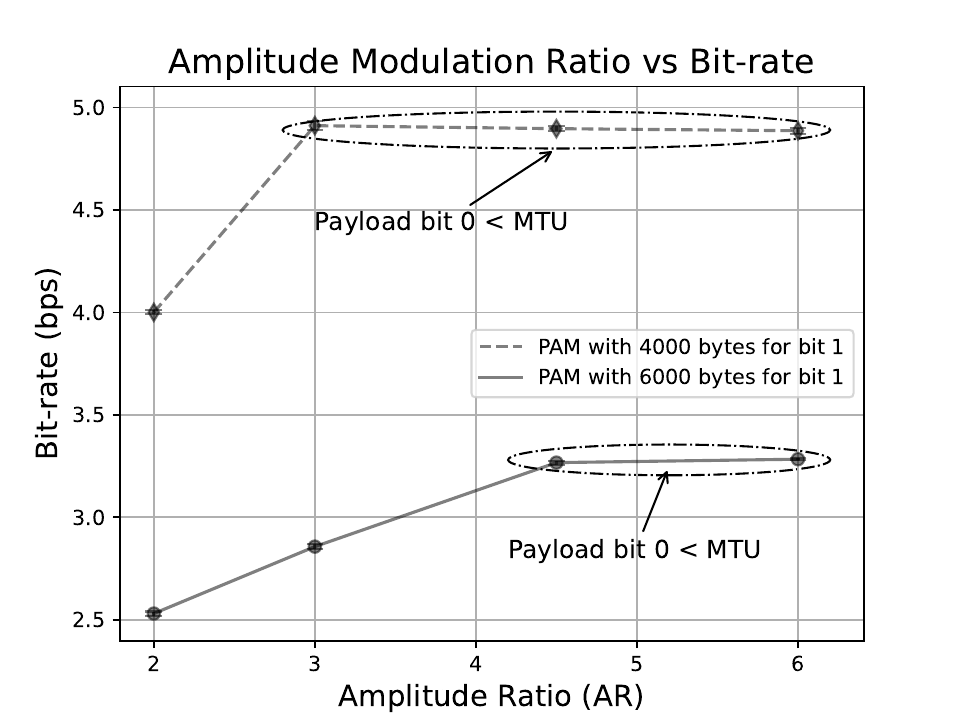}
    \caption{Bit-rate evaluation at varying amplitude modulation depth, i.e., \ac{ar}.}
    \label{fig:AR}
\end{figure}

To evaluate the covert channel's capacity, we conducted measurements to determine its maximum bit-rate. Initially, we set a fixed payload size for transmitting bit `1' ($b1$) and explored various scenarios using smaller payloads to represent bit `0' ($b0$). This approach allowed us to achieve amplitude modulation with variable amplitude ratios (AR). Figure~\ref{fig:AR} illustrates two cases where b1 is set to $6000$ bytes in one group of tests and $4000$ in another.
The results revealed that the maximum achievable bit-rate is approximately $5$~bps. Interestingly, Figure~\ref{fig:AR} shows that the throughput remains constant when the \ac{ar} is such that the payload size of bit zero is smaller than the \ac{mtu}.

\begin{figure}[!ht]
    \centering
    \includegraphics[width=0.9\columnwidth]{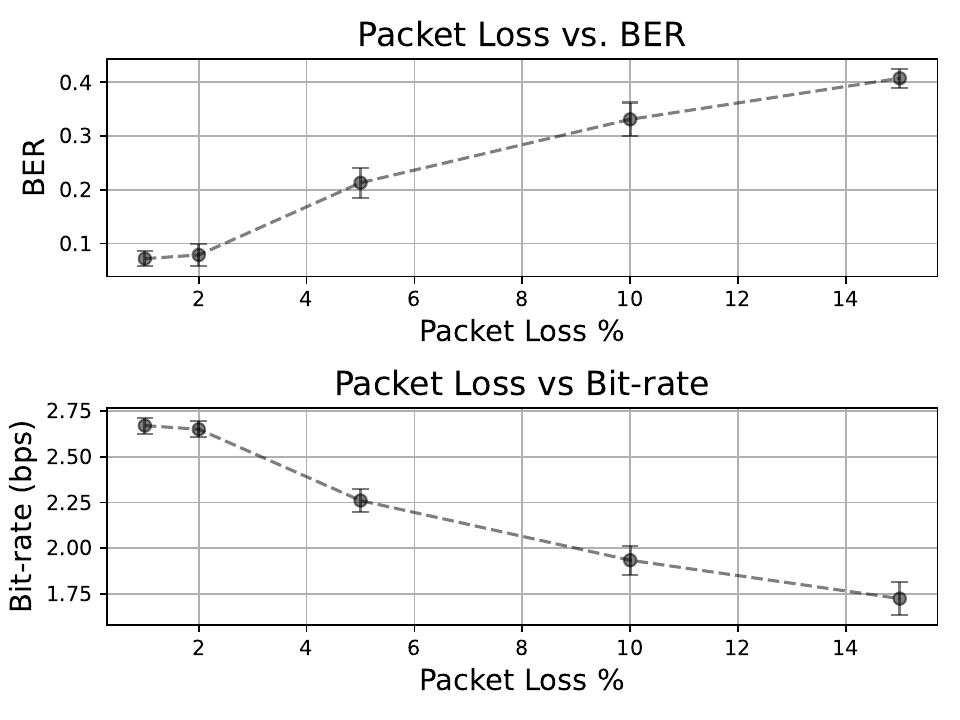}
    \caption{Transmitting a bitstream over the covert channel under packet loss.}
    \label{fig:pktloss}
\end{figure}

\begin{figure}[!ht]
    \centering
    \includegraphics[width=0.9\columnwidth]{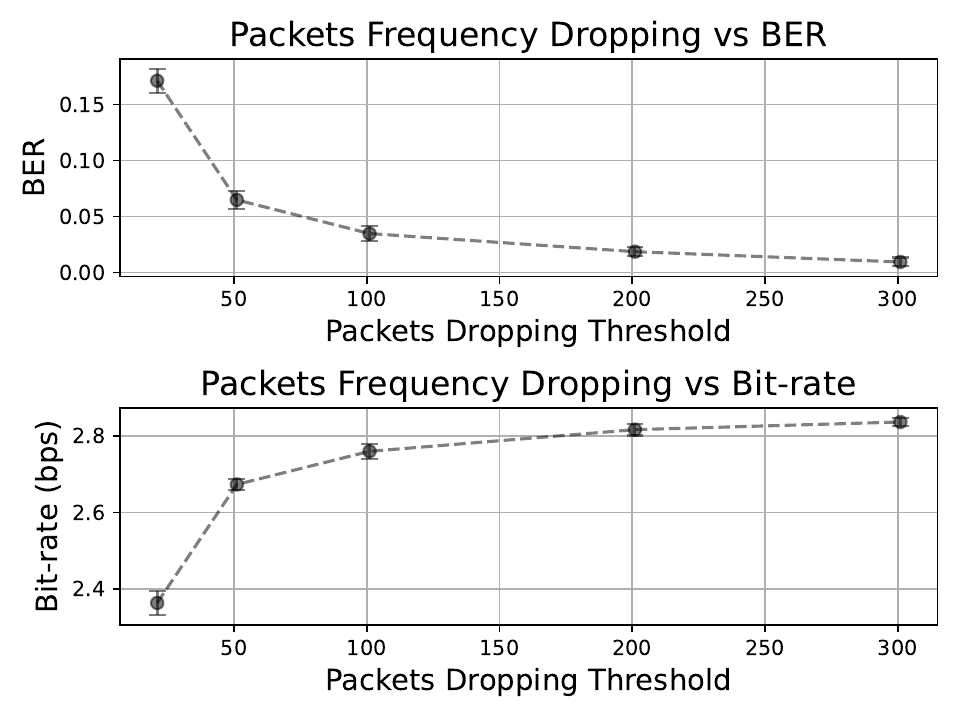}
    \caption{Transmitting a bitstream over the covert channel under packet dropping.}
    \label{fig:freqdrop}
\end{figure}

\begin{figure}[!ht]
    \centering
    \includegraphics[width=0.9\columnwidth]{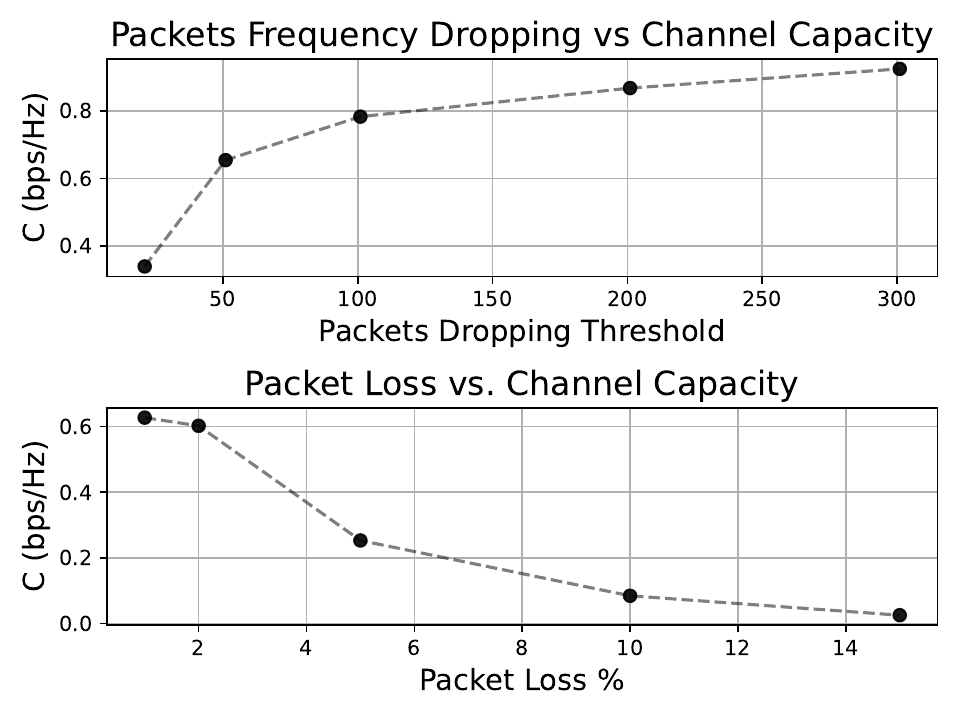}
    \caption{Channel capacity (C) under packets dropping and packet loss.}
    \label{fig:capacity}
\end{figure}

In Figure~\ref{fig:pktloss}, we examined the impact of packet loss on the covert channel using an \ac{pam} scheme with b1 set to $6000$ bytes. Notably, for high packet loss rates, specifically around $5$\%, the \ac{ber} reached approximately $20$\%.
Similarly, Figure~\ref{fig:freqdrop} illustrates the effect of dropping packets periodically, where the threshold denotes the number of packets dropped. For example, with a drop rate of $1$ packet per $50$, the BER was approximately $15$\%.
With the same settings, Figure~\ref{fig:capacity} shows the relation between the channel capacity (C) expressed by \eqref{EQ:CH_C} and packets dropping and packet loss, respectively. With the covert channel proposed here, we achieved a spectral efficiency of $0.9239$~$\sfrac{bps}{Hz}$.

{\color{black}In addition, we conducted further tests
to evaluate the other characteristics of covert channels by considering UDP traffic of up to $10$~Mbps between PC1-1 and PC1-2 and experienced no interference of the covert channel.} We also introduced variations in latency, jitter, and corrupted packets 
(values used for these tests are in Table ~\ref{TAB:PARAM}). Remarkably, none of these operations disrupted the covert channel's transmission. GNS3 allowed us to vary packet loss, frequency drop, latency, jitter, and corruption by operating on the communication link between the two machines.

These results demonstrate the robustness of the covert channel in a typical enterprise network scenario.  Moreover, the channel exhibits a sufficient bit-rate for effective information exfiltration. Its inherent anonymity poses significant challenges for detection since the transmitted bursts exhibit very low throughput, thus avoiding the microburst phenomenon typically mitigated by modern network devices. Additionally, using random payload data and commonly employed UDP services further contribute to the channel's elusive nature, making it difficult to detect.
{\color{black}
\section{Discussion}\label{sec:discussion}
Covert channels have been an area of significant research interest, with various techniques being proposed over the years. The proposed attack offers a novel approach in this domain. As described in the previous sections, the throughput achieved is comparable with other solutions proposed in the literature. In addition, the simplicity of the attack, based on bit-rate modulation, could prospectively make it applicable even in contexts where devices with low computational capabilities operate.

\parag{Comparison with other schemes} While there is  a limited number of studies in the literature that employ a similar scheme to our attack, it remains appropriate to compare our performance with those existing contributions.
In \cite{2012:high-speed_covert-ch}, the authors introduced a covert channel mechanism in a virtualized environment that posed serious security risks in the cloud. Their method achieved a throughput of 100 bps on average. 
In \cite{2022:network_iot_covert_channel}, authors create a unique covert timing channel to exfiltrate data utilizing Internet of Things (IoT) devices. The TCP/IP throughput of the Stealth mode is 4.61 bps, while the ZigBee throughput is $3.90$~bps. 
Recently, Hou~\textit{et al.} \cite{2022:covert_ch_LoRa} published an innovative approach, exploiting the LoRa physical layer to establish a covert channel with a throughput of $30$~bps. This approach is the closest work to our contribution. The authors implemented the covert channel by modulating the amplitude of the LoRA radio signal.

\parag{Mitigation}The detection of the attack we propose is challenging without prior knowledge. Nevertheless, if the network-controlling team possesses specific information, it is plausible that by analyzing network traffic variations (such as observing the network traffic envelope), they could potentially detect an amplitude modulation and trigger an alert.

\parag{Ethical aspects} Our research aims to signal a new class of vulnerabilities within network traffic, thereby hardening defences against exploiting such channels to leak sensitive data by adverse entities. To effectively identify and cancel bit-rate modulation covert channels, network devices must incorporate advanced monitoring capabilities to monitor and respond to anomalous throughput fluctuations.}

\section{Conclusions}\label{sec:conclusions}
We have presented a novel covert timing channel that exploits bit-rate modulation as a new attack vector on enterprise networks. 
{\color{black}An advantage of this technique is that it uses a light and simple algorithm to implement the sender and receiver, making it suitable {\color{black}also} for hacking those resource-limited devices belonging to two \acp{lan} communicating via an enterprise network.}

We conducted extensive experiments in a controlled cyber range environment to validate our approach, allowing us to manipulate network parameters and thoroughly evaluate our model. The results demonstrated the effectiveness of our covert channel, achieving a data transmission rate of $5$~bps and a channel capacity up to $0.9239$~$\sfrac{bps}{Hz}$.  
Although the bit-rate may seem low, let us remember that it is a typical value for a covert channel and sufficient to exfiltrate sensitive data before being discovered.
Furthermore, our solution shows resilience against common network challenges such as jitter, latency, and coexistence with legitimate traffic and maintains complete anonymity.
Through a comprehensive understanding of covert channel attacks, we can defend our networks, preserving the integrity and privacy of data and, ultimately, fostering trust in our digital world.

{\color{black}
\section*{Acknowledgments } \label{SEC:ACK}
The authors thank Niccolò Maggioni for his insightful comments.
This work was partly supported by Consorzio Interuniversitario Nazionale per l'Informatica (CINI)  through a Research Project under Grant CA 01/2021 a.i. 2 and by project SERICS (PE00000014) under the NRRP MUR program funded by the EU - NGEU.}

%
%
%
\bibliographystyle{splncs04}
\bibliography{biblio}

\begin{thebibliography}{10}
\providecommand{\url}[1]{\texttt{#1}}
\providecommand{\urlprefix}{URL }
\providecommand{\doi}[1]{https://doi.org/#1}

\bibitem{gns3}
{Graphical Network Simulator 3 (GNS3)}, \url{https://www.gns3.com/}

\bibitem{scapy}
{Scapy}, \url{ https://scapy.net/}

\bibitem{tcpdump}
{TCPdump}, \url{https://www.tcpdump.org/}

\bibitem{vyos}
{VyOS}, \url{https://vyos.io/}

\bibitem{2021:cyberphy_covert_ch}
Abdelwahab, A., Lucia, W., Youssef, A.: Covert channels in cyber-physical
  systems. IEEE Control Systems Letters  \textbf{5}(4),  1273--1278 (2021).
  \doi{10.1109/LCSYS.2020.3033059}

\bibitem{2022:maritime-covert-channel-esorics}
Amro, A., Gkioulos, V.: {From Click To Sink: Utilizing AIS For Command And
  Control In Maritime Cyber Attacks}. In: Computer Security – ESORICS 2022:
  27th European Symposium on Research in Computer Security, Copenhagen,
  Denmark, September 26–30, 2022, Proceedings, Part III. p. 535–553.
  Springer-Verlag, Berlin, Heidelberg (2022).
  \doi{10.1007/978-3-031-17143-7\_26}

\bibitem{behringer2005mpls}
Behringer, M.H., Morrow, M.: Mpls VPN security. Cisco Press (2005)

\bibitem{2018:new_trends_info_hiding}
Cabaj, K., Caviglione, L., Mazurczyk, W., Wendzel, S., Woodward, A., Zander,
  S.: The new threats of information hiding: The road ahead. IT Professional
  \textbf{20}(3),  31--39 (2018). \doi{10.1109/MITP.2018.032501746}

\bibitem{CISCO2007MPLSDCN}
{Cisco Systems}: {MPLS in the DCN} (2007),
  \url{https://www.cisco.com/c/en/us/td/docs/ios/solutions_docs/telco_dcn/Book/telco5.html}

\bibitem{CISCO2020MPLSVPN}
{Cisco Systems}: {Configuring a Basic MPLS VPN} (2020),
  \url{https://www.cisco.com/c/en/us/support/docs/multiprotocol-label-switching-mpls/mpls/13733-mpls-vpn-basic.html}

\bibitem{2022:soderi-covert-channel-fl}
Costa, G., Pinelli, F., Soderi, S., Tolomei, G.: Turning federated learning
  systems into covert channels. IEEE Access  \textbf{10},  130642--130656
  (2022). \doi{10.1109/ACCESS.2022.3229124}

\bibitem{DeGhein2016MPLS}
Ghein, L.D.: {MPLS Fundamentals}. Cisco Press (2016)

\bibitem{2020:6g-horizons}
Gui, G., Liu, M., Tang, F., Kato, N., Adachi, F.: {6G: Opening New Horizons for
  Integration of Comfort, Security, and Intelligence}. IEEE Wireless
  Communications  \textbf{27}(5),  126--132 (2020).
  \doi{10.1109/MWC.001.1900516}

\bibitem{2019:guri_HOTSPOT}
Guri, M.: {HOTSPOT: Crossing the Air-Gap Between Isolated PCs and Nearby
  Smartphones Using Temperature}. In: 2019 European Intelligence and Security
  Informatics Conference (EISIC). pp. 94--100 (2019).
  \doi{10.1109/EISIC49498.2019.9108874}

\bibitem{2020:guri_acoustic_covert_ch}
Guri, M.: {CD-LEAK: Leaking Secrets from Audioless Air-Gapped Computers Using
  Covert Acoustic Signals from CD/DVD Drives}. In: 2020 IEEE 44th Annual
  Computers, Software, and Applications Conference (COMPSAC). pp. 808--816
  (2020). \doi{10.1109/COMPSAC48688.2020.0-163}

\bibitem{2021:guri_MAGNETO_covert_channel}
Guri, M.: {MAGNETO: Covert channel between air-gapped systems and nearby
  smartphones via CPU-generated magnetic fields}. Future Generation Computer
  Systems  \textbf{115},  115--125 (2021).
  \doi{https://doi.org/10.1016/j.future.2020.08.045},
  \url{https://www.sciencedirect.com/science/article/pii/S0167739X2030916X}

\bibitem{2022:etherled-covert-channel}
Guri, M.: {ETHERLED: Sending Covert Morse Signals from Air-Gapped Devices via
  Network Card (NIC) LEDs}. In: 2022 IEEE International Conference on Cyber
  Security and Resilience (CSR). pp. 163--170 (2022).
  \doi{10.1109/CSR54599.2022.9850284}

\bibitem{2022:nf-covert-channel}
Guri, M.: Near field air-gap covert channel attack. In: 2022 IEEE International
  Conference on Trust, Security and Privacy in Computing and Communications
  (TrustCom). pp. 490--497 (2022). \doi{10.1109/TrustCom56396.2022.00074}

\bibitem{2023:guri_powersupply_covert_ch}
Guri, M.: {POWER-SUPPLaY: Leaking Sensitive Data From Air-Gapped, Audio-Gapped
  Systems by Turning the Power Supplies into Speakers}. IEEE Transactions on
  Dependable and Secure Computing  \textbf{20}(1),  313--330 (2023).
  \doi{10.1109/TDSC.2021.3133406}

\bibitem{2015:guri_thermal_covert_ch}
Guri, M., Monitz, M., Mirski, Y., Elovici, Y.: Bitwhisper: Covert signaling
  channel between air-gapped computers using thermal manipulations. In: 2015
  IEEE 28th Computer Security Foundations Symposium. pp. 276--289 (2015).
  \doi{10.1109/CSF.2015.26}

\bibitem{2018:guri_MOSQUITO_cpvert_ch}
Guri, M., Solewicz, Y., Elovici, Y.: Mosquito: Covert ultrasonic transmissions
  between two air-gapped computers using speaker-to-speaker communication. In:
  2018 IEEE Conference on Dependable and Secure Computing (DSC). pp.~1--8
  (2018). \doi{10.1109/DESEC.2018.8625124}

\bibitem{2022:network_iot_covert_channel}
Harris, K., Henry, W., Dill, R.: {A Network-based IoT Covert Channel}. In: 2022
  4th International Conference on Computer Communication and the Internet
  (ICCCI). pp. 91--99 (2022). \doi{10.1109/ICCCI55554.2022.9850247}

\bibitem{2022:covert_ch_LoRa}
Hou, N., Xia, X., Zheng, Y.: {CloakLoRa: A Covert Channel Over LoRa PHY}.
  IEEE/ACM Transactions on Networking pp. 1--14 (2022).
  \doi{10.1109/TNET.2022.3209255}

\bibitem{2009:detection-SS-wm-nordsec}
Jia, W., Tso, F.P., Ling, Z., Fu, X., Xuan, D., Yu, W.: Blind detection of
  spread spectrum flow watermarks. In: IEEE INFOCOM 2009. pp. 2195--2203
  (2009). \doi{10.1109/INFCOM.2009.5062144}

\bibitem{2018:bridges_towards_katz}
{Katz}, M., {Matinmikko-Blue}, M., {Latva-Aho}, M.: {6Genesis Flagship Program:
  Building the Bridges Towards 6G-Enabled Wireless Smart Society and
  Ecosystem}. In: 2018 IEEE 10th Latin-American Conference on Communications
  (LATINCOM). pp.~1--9 (Nov 2018). \doi{10.1109/LATINCOM.2018.8613209}

\bibitem{2013:real-story-stuxnet}
Kushner, D.: The real story of stuxnet. IEEE Spectrum  \textbf{50}(3),  48--53
  (2013). \doi{10.1109/MSPEC.2013.6471059}

\bibitem{1973:lampson_covert_ch}
Lampson, B.W.: A note on the confinement problem. Communications of the ACM
  \textbf{16}(10),  613--615 (Oct 1973),
  \url{https://doi.acm.org/10.1145/362375.362389}

\bibitem{2014:PHY_covert_ch}
Lee, K.S., Wang, H., Weatherspoon, H.: {PHY} covert channels: Can you see the
  idles? In: 11th USENIX Symposium on Networked Systems Design and
  Implementation (NSDI 14). pp. 173--185. USENIX Association, Seattle, WA (Apr
  2014),
  \url{https://www.usenix.org/conference/nsdi14/technical-sessions/presentation/lee}

\bibitem{2008:tcp_covert_timing_channel}
Luo, X., Chan, E.W.W., Chang, R.K.C.: Tcp covert timing channels: Design and
  detection. In: 2008 IEEE International Conference on Dependable Systems and
  Networks With FTCS and DCC (DSN). pp. 420--429 (2008).
  \doi{10.1109/DSN.2008.4630112}

\bibitem{2011:timinig-wm-nordsec}
Luo, X., Zhou, P., Zhang, J., Perdisci, R., Lee, W., Chang, R.K.C.: Exposing
  invisible timing-based traffic watermarks with backlit. In: Proceedings of
  the 27th Annual Computer Security Applications Conference. p. 197–206.
  ACSAC '11, Association for Computing Machinery, New York, NY, USA (2011).
  \doi{10.1145/2076732.2076760}, \url{https://doi.org/10.1145/2076732.2076760}

\bibitem{Proakis2007digital}
Massoud~Salehi, P., Proakis, J.: Digital Communications 5th Edition.
  McGraw-Hill Education (2007), {ISBN:} 9780072957167

\bibitem{2015:info_hiding_malware_detection}
Mazurczyk, W., Caviglione, L.: Information hiding as a challenge for malware
  detection. IEEE Security \& Privacy  \textbf{13}(2),  89--93 (2015).
  \doi{10.1109/MSP.2015.33}

\bibitem{2013:timing_covert_channel_retx}
Mazurczyk, W., Smolarczyk, M., Szczypiorski, K.: On information hiding in
  retransmissions. Telecommunication Systems  \textbf{52}(2),  1113 – 1121
  (2013). \doi{10.1007/s11235-011-9617-y}

\bibitem{2009:storage_covert_ch_fragments}
Mazurczyk, W., Szczypiorski, K.: Steganography in handling oversized ip
  packets. In: 2009 International Conference on Multimedia Information
  Networking and Security. vol.~1, pp. 559--564 (2009).
  \doi{10.1109/MINES.2009.246}

\bibitem{2009:stega_storage_covert_channel}
Mazurczyk, W., Szczypiorski, K.: Steganography in handling oversized ip
  packets. vol.~1, p. 559 – 564 (2009). \doi{10.1109/MINES.2009.246}

\bibitem{2017:information_hiding}
Mazurczyk, W., Wendzel, S.: Information hiding: Challenges for forensic
  experts. Commun. ACM  \textbf{61}(1),  86–94 (dec 2017).
  \doi{10.1145/3158416}, \url{https://doi.org/10.1145/3158416}

\bibitem{2022:covert_ch_detection}
Ondov, A., Helebrandt, P.: Covert channel detection methods. In: 2022 20th
  International Conference on Emerging eLearning Technologies and Applications
  (ICETA). pp. 491--496 (2022). \doi{10.1109/ICETA57911.2022.9974878}

\bibitem{2019:router-covert-channel-usenix}
Ovadya, A., Ogen, R., Mallah, Y., Gilboa, N., Oren, Y.: Cross-router covert
  channels. In: Proceedings of the 13th USENIX Conference on Offensive
  Technologies. p.~2. WOOT'19, USENIX Association, USA (2019)

\bibitem{2014:soderi_fingerprint}
Soderi, S., Dainelli, G., Iinatti, J., Hämäläinen, M.: Signal fingerprinting
  in cognitive wireless networks. In: 2014 9th International Conference on
  Cognitive Radio Oriented Wireless Networks and Communications (CROWNCOM). pp.
  266--270 (2014). \doi{10.4108/icst.crowncom.2014.255374}

\bibitem{2018:strom_mitre}
Strom, B.E., Applebaum, A., Miller, D.P., Nickels, K.C., Pennington, A.G.,
  Thomas, C.B.: {MITRE ATT\&CK: Design and philosophy}. In: Technical report.
  The MITRE Corporation (2018)

\bibitem{2016:APT}
Ussath, M., Jaeger, D., Cheng, F., Meinel, C.: Advanced persistent threats:
  Behind the scenes. In: 2016 Annual Conference on Information Science and
  Systems (CISS). pp. 181--186 (2016). \doi{10.1109/CISS.2016.7460498}

\bibitem{2015:network-covert-ch-nordsec}
Wendzel, S., Zander, S., Fechner, B., Herdin, C.: Pattern-based survey and
  categorization of network covert channel techniques. ACM Comput. Surv.
  \textbf{47}(3) (apr 2015). \doi{10.1145/2684195},
  \url{https://doi.org/10.1145/2684195}

\bibitem{2022:TCAN_conti}
Ying, X., Bernieri, G., Conti, M., Bushnell, L., Poovendran, R.: Covert
  channel-based transmitter authentication in controller area networks. IEEE
  Transactions on Dependable and Secure Computing  \textbf{19}(4),  2665--2679
  (2022). \doi{10.1109/TDSC.2021.3068213}

\bibitem{2012:high-speed_covert-ch}
Zhenyu, W., Zhang, X., Wang, H.: Whispers in the hyper-space: high-speed covert
  channel attacks in the cloud. In: USENIX Security symposium. pp. 159--173
  (2012)

\end{thebibliography}

\end{document}